\documentclass[10pt,letterpaper]{article}
\usepackage[utf8]{inputenc}
\usepackage[english]{babel}
\usepackage{amsmath}
\usepackage{amsfonts}
\usepackage{amssymb}
\usepackage{graphicx}
\usepackage{graphics}
\usepackage[left=2cm,right=2cm,top=2cm,bottom=2cm]{geometry}
\usepackage{multicol}
\usepackage{multirow}
\usepackage{color}
\usepackage{hyperref} 

\begin{document}
\title{\textbf{\huge{}Origin of Information Encoding in Nucleic Acids through a Dissipation-Replication Relation}}

\author{Juli\'an Mej\'ia$^{1}$ and Karo Michaelian$^{2}$}

\maketitle

\begin{center}

\begin{center}
{\small{} [1] Postgrado in Physical Sciences, Instituto de F\'isica, Universidad Nacional Aut\'onoma de M\'exico, Apartado Postal 20-364, M\'exico D. F. 01000, M\'exico.}
\par\end{center}{\small \par}

{\small{} [2] Department of Nuclear Physics and Application of Radiation, Instituto de F\'isica, Universidad Nacional Aut\'onoma de M\'exico, Apartado Postal 20-364, M\'exico D. F. 01000, M\'exico.}
\par\end{center}{\small \par}

\begin{center}
\textbf{\footnotesize{}karo@fisica.unam.mx; julianmejia@ciencias.unam.mx}
\par\end{center}{\footnotesize \par}

\begin{abstract}
Ultraviolet light incident on organic material can initiate its spontaneous dissipative structuring into chromophores which can then catalyze their own replication. This may have been the case for one of the most ancient of all chromophores dissipating the Archean UVC photon flux, the nucleic acids. Under the empirically established imperative of increasing entropy production, nucleic acids with affinity to particular amino acids which foment UVC photon dissipation would have been ``thermodynamically selected'' through this dissipation-replication relation. Indeed, we show here that those amino acids with characteristics most relevant to fomenting UVC photon dissipation are precisely those with greatest affinity to their codons or anticodons. This could provide a physical-chemical mechanism for the accumulation of information in nucleic acids relevant to the dissipation of the externally imposed photon potential. This mechanism could provide a non-equilibrium thermodynamic foundation, based on increasing global entropy production of the biosphere, for the tenants of Darwinian natural selection. We show how this mechanism might have begun operating at the origin of life in the Archean, and how, in fact, it still operates today, albeit indirectly through complex biosynthetic pathways now operating in the visible.

\end{abstract}

\section{{\large{}Introduction}}
An explanation for the origin of the genetic code has been a difficult problem since its resolution is intimately related with the origin of life itself. From the traditional perspective of life as a rather fortuitous enzyme-catalyzed reproductive event, the problem arises in that sufficient information must have somehow been acquired in the incipient genome before such a faithful reproductive event could have taken place. Due to the complexity of such an event, it is highly unlikely that that this information could have been generated randomly. 

A number of theories have attempted to addressed this problem by considering the origin of the association between amino acids and their cognate codons or anticodons. The ``stereochemical theory'' suggests that amino acids became linked with their codons or anticodons based on stereo-chemical affinity\cite{Woese1967,Yarus1998,YarusEtAl2005}, while the ``frozen accident theory'' \cite{Crick1968,Koonin2017} suggests that such links arose purely by chance and once established would have remained so since any changes would have been highly deleterious to protein construction and surely detrimental to the organism. The ``co-evolutionary theory'' \cite{Wong1975} suggests instead that the structure of the codon system is primarily an imprint of the prebiotic pathways of amino-acid formation, which remain recognizable in contemporary enzymic pathways of amino-acid biosynthesis. 

All theories thus far presented only partially address the problem of the origin of the genetic code since, although they provide reasons for the association between particular amino acids and their cognate codons, there is no physical-chemical description of how the specificity of such an association relates to the origin of life, in particular, to enzyme-less reproduction, proliferation and evolution. Carl Woese \cite{Woese1967} recognized this early on and emphasized the probelm, still unresolved, of uncovering the basis of the specifity between amino acids and codons in the genetic code.  

In this paper, from within the frameworks of the stereochemical theory and non-equilibrium thermodynamics, we propose a novel theory for the physical-chemical basis of the specificity of these associations between codons/anticodons and amino acids. The basis we propose is related to the non-equilibrium thermodynamic imperative of the production of entropy for the origin, proliferation, and evolution of the irreversible process known as life. In particular, we consider the dissipation of the Archean solar photon spectrum at Earth's surface as the driver of the origin and evolution of life and show how information related to the efficient dissipation of this solar photon potential could have been the first information programmed into the genome.

Clasical Irreversible Thermodynamic (CIT) theory, formulated by Lars Onsager \cite{Onsager1931a,Onsager1931b,OnsagerMachlup1953} and Ilya 
Prigogine \cite{Prigogine1955} indicates that all irrevesible processes arise, persist, and 
even evolve to dissipate a generalized chemical potential, i.e. to produce entropy. 
Glansdorff and Prigogine have shown how a system over which a generalized chemical potential is imposed can ``self-organize'' into structures (or more correctly, processes) which break symmetry in both space and time \cite{GlansdorffPrigogine1971}. This organization of material is known as "dissipative structuring'' in non-equilibrium thermodynamic language. Biologists are acutely aware of a sub-set of this structuring and refer to it simply as``life". 

Macroscopic dissipative structuring leads to macroscopic processes such as hurricanes, 
winds, ocean currents and the water cylce. However, microscopic dissipative 
structuring can also occur \cite{Michaelian2017}, in which the structuring involves the molecular internal degrees of freedom; isomeric or tautomeric reconfiguration, rotation about covalent bonds, electric polarization or spin orientation in an external electric or magnetic field, electronic excited states, exciplex and excimer formation, etc..  In the case of structuring involving molecular configurational degrees of freedom, the structuring can remain even after the removal of the impressed generalized chemical potential due to strong inter-atomic forces \cite{Michaelian2016}. 

The most important generalized chemical potential (source of free energy) at Earth's surface today, and also at the origin of life at $\sim 3.85$ Ga, is the solar photon potential; the sun's spectrum with respect to the cosmic black-body radiation filling space surrounding Earth. During the Archean, sunlight surpassed all other forms of available energy (hydrothermal vents, lightening, chemical potentials, shock waves, etc.) by at least three orders of  magnitude \cite{FerrisChen1975}. The most important dissipative structures existing on  Earth's surface today are the organic pigments
in water solvent \cite{Michaelian2016} and these are responsible for approximately 63\% of the total entropy production resulting from Earth's interaction with its solar environment \cite{KleidonLorenz2005}. The coupling of the heat of dissipation
of the solar photons in organic pigments to other macroscopic irreversible processes such as ocean and wind currents and the water cycle, accounts for the majority of the rest of the entropy production on Earth \cite{Kleidon2009}. 

Without the pre-existence of complex biosynthetic pathways, the
photon wavelengths that could have bootstrapped life through dissipation at life's origin in the Archean must have involved the long wavelength UVC and UV-B regions, where there is enough free energy to make and break carbon covalent bonds, but not enough to ionize and thereby destroy organic molecules. A long wavelength UVC component of the solar  spectrum indeed penetrated to the surface of Earth throughout the Archean with an integrated intensity of up to 5 W/m$^2$ midday at the equator \cite{Sagan1973,Michaelian2009}.  Evolving life was exposed to this UVC photon potential for at least 1,000 million years (1 Ga) until approximately 2.7 Ga when organisms performing oxygenic photosynthesis became abundant 
enough to overwhelm the natural abiotic oxygen sinks \cite{Kasting1993}. 

Corrborating evidence for the exposure of life to this UVC photon potential, and indeed life's 
thermodynamic preocupation with its dissipation, can be found in the fundamental molecules 
of life (those common to all three domains). Most of these absorb and dissipate this light into heat
with great efficiency \cite{Michaelian2016, MichaelianSimeonov2015}. We have therefore suggested
that the fundamental molecules originated in the Archean as microscopic dissipative structures  to perform this thermodynamic function \cite{Michaelian2017, Michaelian2009, MichaelianSimeonov2015, Michaelian2011}. 

The proliferation  of these pigments over the surface of Earth, a hallmark of biological evolution, can be explained through the autocatalytic nature of these pigments in dissipating the same UVC potential that produced them photochemically \cite{MichaelianSimeonov2015}.  Analagously to what Prigogine \cite{Prigogine1967} has demonstrated using CIT theory for autocatalytic chemical reactions, the concentration of a photochemical product (chromophore or pigment) will increase much beyond its expected equilibrium concentration if the product acts as a catalyst for the dissipation the same impressed photon potential that produced it \cite{Michaelian2013}. For fundamental chromophores of the Archean such as the nucelic acids absorbing in the UVC, this leads to a direct relation between dissipation and replication. Proliferation of these fundamental molecules in this manner over the whole of Earth's surface is thereby driven by thermodynamic imperative of increasing the global entropy production of Earth in its solar environment. Although dissipative structuring of material invariably leads to a reduction in its entropy, that of the system plus environment invariably increases, thereby respecting the second law of thermodynamics.

Since microscopic dissipative structuring can be persistent, i.e. structuring remains even after the removal of the impressed external potential, information concerning the impressed potential and the molecular structuring required for its optimal dissipation becomes programmed into the nucleic acid. The cumulative historical information regarding the external generalized chemical potentials, and the blueprint for construction of the biosynthetic pathways needed for the production of structures (e.g. chromophores and their supporting structures) required for dissipation of those potentials, is today known as the {\it genome}. Such a microscopic mechanism with inherent persistence in structure allows the system to ``evolve'', i.e. to optimally adapt the dissipative structure to changes in the external potential, or to resume dissipation should the external potential return after temporal absence (for example, after the overnight extinction of the solar photon potential). 

The mechanism (to be described below) relating replication with dissipation, under the non-equilibrium thermodynamic imperative of increasing the global entropy production, allows the system to evolve to ever greater efficacy in photon dissipation (by, for example, increasing the areal coverage of pigments over Earth's surface, or increasing the wavelength region of dissipation) often with a corresponding increase in system complexity, eventually arriving at the complex global dissipative structure of today known as the \textit{biosphere} which involves the coupling of biotic with abiotic dissipative processes.

Based on these premises, in this article we develop a theory for the origin of the genetic code and for its coevolution with the fundamental molecules of life. The theory assigns entropy production through photon dissipation in the early Sun-Earth system as the driver for the initiation of this process, and the increase in entropy production as the criterion for selection of new coupling of distinct irreversible processes involving microscopic dissipative structures. The goal of this article is to describe a plausible mechanism for the initiation of information accumulation in the ancient genome through the action of a proposed dissipation-replication relation existing under the imposed UVC photon potential prevalent at the very beginnings of life on Earth. Before describing this mechanisms in detail, we first 
discuss some of the salient properties of the fundamental molecules of life which provide evidence for, and are crucial to, the viability of such a mechanism.

\section{Salient Properties of Amino Acid, Nucleic Acids, and their Complexes}

\subsection{Amino acids and Nucleic acids absorb and dissipate in the UVC region.}
\begin{figure}[htbp]
\begin{center}
\includegraphics[angle=90,width=530px,height=280px]{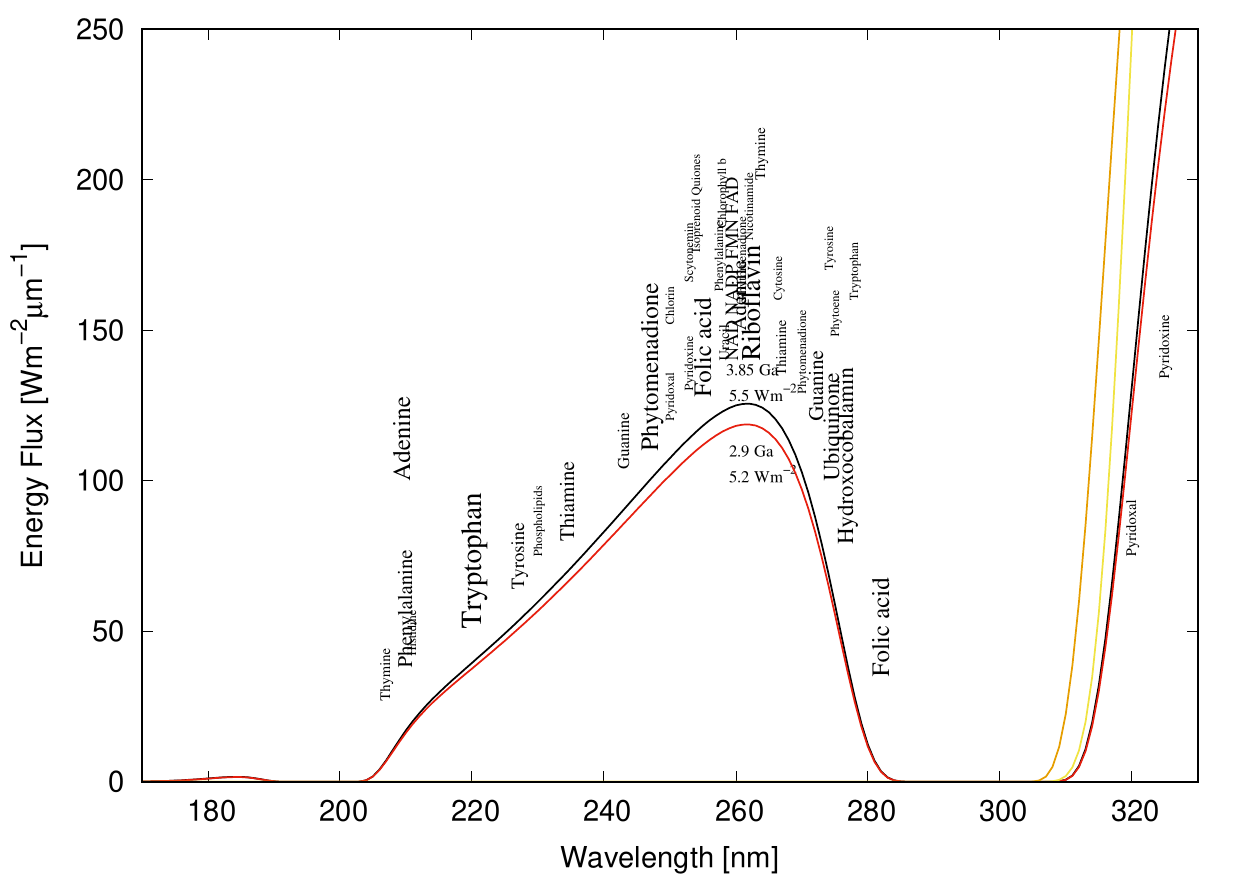} 
\end{center}
\caption{The wavelengths of maximum absorption of many of the fundamental molecules of life (common to all three domains), including the the nucleobases, aromatic and charged amino acids, cofactors, vitamins, and lipids coincide with the predicted solar spectrum at Earth's surface in the UVC at the time of the origin of life at 3.85 Ga (black line) \cite{Sagan1973} and until at least 2.9 Ga (red line). The font size of the pigment name roughly indicates the relative size of its molar extinction coefficient. Image credit: Adapted from \cite{MichaelianSimeonov2015}.}
\label{FundamentalMolecules}
\end{figure}

The fundamental molecules of life (those molecules common to all three domains of life) and their associations in complexes are examples of microscopic self-organized dissipative structures which formed under the Archean UVC photon potential \cite{Michaelian2009, MichaelianSimeonov2015, Michaelian2011, Michaelian2013, MichaelianOliver2011}. Evidence for this can be found in the strong absorption cross sections of the fundamental molecules, with absorption maxima lying exactly within the predicted UVC window of Earth's atmosphere during the Archean \cite{Sagan1973} (see figure \ref{FundamentalMolecules}), and in the existence of inherent conical intersections doting these same molecules with extraordinarily rapid non-radiative excited state decay\cite{Michaelian2016, MiddletonEtAl2009}. 

The nucleobases, and also their polymerization into single and double strand  RNA and DNA, absorb and dissipate UVC light and can be formed from simpler precursor molecules such as hydrogen cyanide HCN in water by the very same UVC photons that they eventually dissipate with such efficacy \cite{Michaelian2016,Michaelian2017}. An example of such microscopic dissipative structuring under UVC light is the generic photochemical pathway to the purine nucleobases first discovered by Ferris and Orgel in 1966 \cite{FerrisOrgel1966}. Although the particular photochemical reaction pathways to the multitude of other fundamental molecules (UVC pigments) still remain to be discovered, the existence of microscopic dissipative structuring is evidenced, for example, by the large quantity of organic pigment molecules found throughout the cosmos wherever UV light is available \cite{MichaelianSimeonov2017}. Today, a similar situation exists on Earth, with pigments dissipating in the near UV and visible which are constructed through more complex biosynthetic pathways but remain, nonetheless, microscopic dissipative structures dissipating the same visible light that provided the free energy for their production.

\subsection{Amino acid affinity to codons (or anti-codons) of DNA and RNA.}
\label{sec:Affinity}
Stereochemical theories for the origin of the genetic code propose that chemical affinity between amino acids and nucleic acids was the historical basis for the present association between codons and/or anticodons and cognate amino acids \cite{Crick1966,Woese1966,ToulmeEtAl1974, Orgel1974, LaceyWeber1976, WeberLacey1978, HendryWitham1979, HendryEtAl1981b, Alberti1997, KnightAndLandweber1998, Yarus1998, MajerfeldEtAl2005, YarusEtAl2005,  Serganov2008, YarusEtAl2009}. These works present sufficient evidence to suggest that, at some early stage in the evolution of life, amino acids were directly associated with specific polynucleotides which later evolved into the genetic code.  

The coevolution theory of the genetic code \cite{Woese1966, Wong1975} suggests that an initial code, containing a set of codons with affinity for primordial prebiotic amino acids, was later extended to code for amino acids which could be synthesized through emerging biosynthetic pathways. In addition, this theory suggests that the amino acid to nucleic acid coding has changed over the evolutionary history of the code; for example, ancestral triplets escaping from amino acid binding sites to acquire new functions in a more modern translation apparatus as codons and anticodons for new amino acids which were able to be biosynthesized.

Although evidence for the chemical association of all twenty common amino acids with RNA or DNA exists, not all associations are of the same character or strength. There are eight distinct types of associations between amino acids and nucleotides or their polymerizations into nuclei acids; 1) the $\pi$-cation bond\cite{YarusEtAl2009, Dougherty2007}, 2) $\pi$-stacking of aromatic structures \cite{KamiichiEtAl1986}, 3) Coulomb charge interaction \cite{ADEtAL2008}, 4) hydrophobic agregation \cite{WeberLacey1978}, 5) Van der Waals interaction, 6) hydrogen bonding \cite{GoviEtAl1985}, 7) stereochemical (for example, substitution of nucleotides by amino acids of similar structure \cite{HendryEtAl1981b, HendryEtAl1981a}), and 8) the possibility of synthesis of amino acids from acids $\alpha$-ketos bounded to dinucleotides \cite{CopleyEtAl2005}.

Of these eight forms of association, four are non-specific, meaning that
the amino acids do not bind to any particular sequence along the single or double
strand DNA or RNA. However, the other four types of associations; the $\pi$-cation bonding, hydrogen bonding, stereochemical affinity and amino acid synthesis from
$\alpha$-keto, are specific, i.e. bind the amino acids almost unequivocally to either their cognate codons or anticodons.

One of the non-specific associations between nucleic acids and amino acids is due to the aliphatic nature of the side chain of the amino acids which can attach themselves, for example within the grooves, through van der Waals and hydrophobic interactions \cite{YarusEtAl2005,Serganov2008}. Another non-specific interaction is that between aromatic amino acids and nucleobases due to stacking interactions that occur when amino acid side chains contain charged sites like indole rings or aromatic rings, for example, Trp, Tyr, Phe and His \cite{KamiichiEtAl1986}. Due to a favorable contribution from hydrophobic effects, stacking assumes an even greater importance in aqueous solution \cite{GoviEtAl1985}. In addition, when the size of the indole or aromatic ring of the amino acid is similar to that of purine base, for example tryptophan, stacking may involve only one strand of the double helix RNA or DNA \cite{ToulmeEtAl1974}.

The types of specific interactions between aromatic amino acids and RNA or DNA are also varied. For example, as a result of the separation of the $\pi$-electrons from the nuclear charges in aromatic molecules, even if the electronic distribution is symmetric, a quadrapole electric moment arises. This allows the aromatic rings to interact as polar elements and form bonds such as $\pi$-cation. This distribution of charge in the aromatic ring confers high specificity to the bond produced between the aromatic amino acids and their cognate codons \cite{YarusEtAl2009, Dougherty2007}.

Possible recognition schemes for amino acids with charged residues are limited\cite{GoviEtAl1985, HendryEtAl1981a}. However, these amino acids can be bound to RNA or DNA because the $\alpha$-amino and $\alpha$-carboxyl groups provide good complements for hydrogen bond receptors and donors \cite{YarusEtAl2009, GoviEtAl1985, HendryEtAl1981a}. Amino acids which can also form a pair of hydrogen bonds with particular bases of nucleic acids are Asp, Glu, Asn and Gln \cite{ HendryEtAl1981a}. 

An interesting example of specific stereochemical affinity between nucleotides and amino acids is related to their structural similarity. In fact, a nitrogenous base of DNA or RNA can be replaced by an amino acid of similar size while maintaining the structural integrity of the nucleic acid. The nucleobase which can be replaced correlates strongly with the second base of the cognate codon\cite{HendryEtAl1981b}.

A hyrdrophobic association has been established between all 20 common amino acids used by life and their codons or anti-codons.  Codons having U as the second base have been associated with the most hydrophobic amino acids, and those having A as the second base are associated with the most hydrophilic amino acids \cite{CopleyEtAl2005}. Some of the found associations are highly specific and may have given rise to assignations for the homocodonic amino acids (Phe, Pro, Gly, Lys) \cite{WeberLacey1978, HendryEtAl1981a, FoxEtAl1971}.

Within the framework of the thermodynamic dissipation theory of the origin of life, the hydrophobic properties of amino acids may have been of great importance since they would have permitted the amino acid-nucleic acid complex to remain at the ocean surface where they would have been exposed to the greatest solar photon potential for fomenting the dissipation-replication relation by promoting photochemical reactions and a photon-induced RNA or DNA denaturing mechanism \cite{MichaelianSantillan2014} described in the following section. 

An example of how photochemical reactions could also have played an important role in the later synthesis of amino acids can be found in the proposal that two bonded nitrogenous bases could have acted as catalysts for the synthesis of simple amino acid synthesis from $\alpha$-keto acids \cite{CopleyEtAl2005}. This could  satisfactorily explain the hydrophobic properties of the simple amino acids and, at the same time, their affinity to their cognate codons (or at least the first two bases of the codon, a ``di-codon''). The hydrophobic property would have allowed the complex to remain at the ocean surface where the UV flux was high and this hydrophobicity is inherited to the synthesized amino acid. Dinucleotides might thereby catalyze reactions required for the synthesis of amino acids by; providing free energy available through photon capture within the nucleotides \cite{Michaelian2017}, orientation and polarization of reactants through hydrogen bonding interactions, use of functional groups as nucleophiles or general bases, attachment of cofactors such as NADH or prebiotic equivalents, use of phosphate groups as an acid or a base catalyst, use of Mg$^{+2}$ ions coordinated to phosphate groups as Lewis acid catalysts, and use of Mg$^{+2}$-coordinated hydroxide ions as nucleophiles or general base catalysts \cite{YasuomiEtAl2004, LipfertEtAl2008, CassanoEtAL2004, NakanoEtAl2000}. UVC light  interacting with nucleotides thus probably played a fundamental role in the early synthesis of these amino acids at some point during the Archean.
 
In some remaining cases it is still not possible to specify or quantify the nature of the interaction that exists between the amino acids and the codons that encode them (e.g. glutamine is very difficult to isolate on RNAs \cite{YarusEtAl2009}) or the case of cysteine or metionine in which the affinity to RNA atamers has been observed by the Systematic Evolution of Ligands by Exponential enrichment (SELEX) method but seems to be non specific \cite{StoltenburgEtAl2007, GebhardtEtAl2000, BurkeAndGold1997}. 

In summary, there exist an intrinsic and specific association between many of the amino acids and their cognate codons or anticodons. Except for the proposal of Copley et al. \cite{CopleyEtAl2005} for the synthesis of the amino acids employing catalytic dinucleotides, and the direct RNA template (DRT) theory for protein synthesis proposed by Yarus \cite{Yarus1998, YarusEtAl2009}, the authors are not aware of any other theory for this specific association. The rest of this article describes our proposal for the chemical-physical basis of this specific association related to UVC photon dissipation which may be more relevant to the very beginnings of the origin of life, before peptides or biosynthesis of amino acids were required.

\section{{\large{} Ultraviolet and Temperature Assisted Replication (UVTAR)}}

This section describes the proposed mechanism by which replication of the RNA/DNA-amino acid complex can be associated with dissipation, a prerequisite for any irreversible process and, specifically for the accumulation of information in the genome through a type of thermodynamic selection summarized below and described in detail in references \cite{Michaelian2016, Michaelian2011}.
 
The temperature of approximately 80-85 $^\circ$C of the ocean surface during the early Archean \cite{LoweTice2004}, when it is generally assumed that life arose, is strikingly similar to the short strand DNA melting temperatures, the temperature at which 50\% of double strand DNA is denatured into single strands, in water at neutral pH and present ocean salinity. This provides an important clue as to the possible nature of enzymeless reproduction at the origin of life.

The temperature at Earth's poles would have been colder, perhaps closer to the denaturing temperature of RNA under similar salt and pH ocean conditions (40 to 50 $^\circ$C), while the temperature at the equator would have been warmer. There would also have been a temperature profile with depth and a diurnal variation of temperature in the ocean microlayer similar to that of today, but of higher absolute temperature \cite{Michaelian2016}. 

The Earth was cooling gradually as the greenhouse gas CO$_2$ was being consumed in silicate carbonates formed through erosion of the newly forming continents \cite{LoweTice2004}, and once the local surface temperature fell below their respective denaturing temperatures, DNA and RNA single strand segments (having been formed as microscopic dissipative structures through autocatalytic photochemical and polymerization routes \cite{Michaelian2016, Michaelian2017}) would eventually find and hydrogen bond with short complementary segments and would then normally be unable to separate again, thus excluding the possibility of template reproduction. However, through absorption by RNA and DNA of UVC light during the day and by dissipating this light into heat deposited locally at the site of the molecule, plus the absorption of solar infrared light on the ocean surface, the local temperature at the site of the molecule may have been raised sufficiently, and for a time sufficiently long, for denaturing to occur, a process that we have called photon-induced denaturing \cite{Michaelian2017, Michaelian2011}.

\begin{figure}[htb]
\begin{center}
\includegraphics{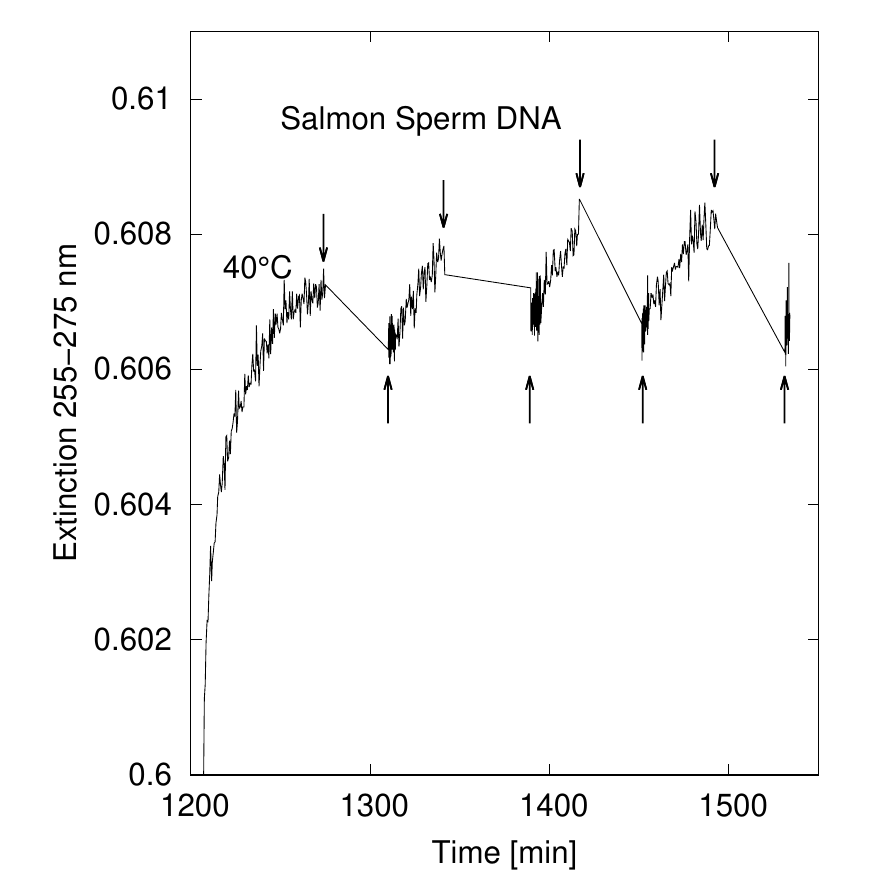}
\end{center}
\caption{Experimental demonstration of UVC light denaturing of Salmon sperm DNA of average length 100 Kbp in pure water (without salt). The graph plots the extinction of the UVC light (due mainly to absorption with a small amount of scattering) by DNA in the wavelength range 255 to 265 nm against time as the UVC light was cycled on and off. The temperature of the bath was raised over a time period of 20 minutes to 40 $^\circ$C (lower arrow in red) and later maintained at this value ($\pm$0.01 $^\circ$C) for the duration of the experiment. The arrows pointing downward mark the time at which the UVC light was blocked from reaching the sample by a shutter and the arrows pointing upwards mark the times at which the light was allowed on sample by removing the shutter. It can be seen that while UVC light is on sample, the extinction increases gradually (after 1/2 hour to about 0.3\% of the differential absorption - between completely denatured and completely natured) due to the hypochromicity arising from UVC light-induced denaturing. While the light was blocked from the sample, the segments renatured. The amount of denaturing depends on the intensity of the UVC light and the temperature of the bath. Image credit: Michaelian and Santillan \cite{MichaelianSantillan2014}.}
\label{Expt}
\end{figure}

This enzymeless denaturing under UVC light is not hypothetical, we have measured it quantitatively for DNA \cite{MichaelianSantillan2014} and it is operative even for water temperatures well below nominal DNA melting temperatures, albeit with decreasing efficiency as the solvent temperature is lowered. In reference \cite{Michaelian2016} we have postulated several UVC induced mechanisms for breaking the hydrogen bonds between RNA or DNA strands. Our experiments have also shown that UVC light-induced denaturing of a DNA oligo is reversible, i.e. most damage inflicted on the molecule by the UVC light is reversible and the molecule can renature again without difficulty (see figure \ref{Expt}).

\begin{figure}[htb]
\begin{center}
\includegraphics[width=460px,height=330px]{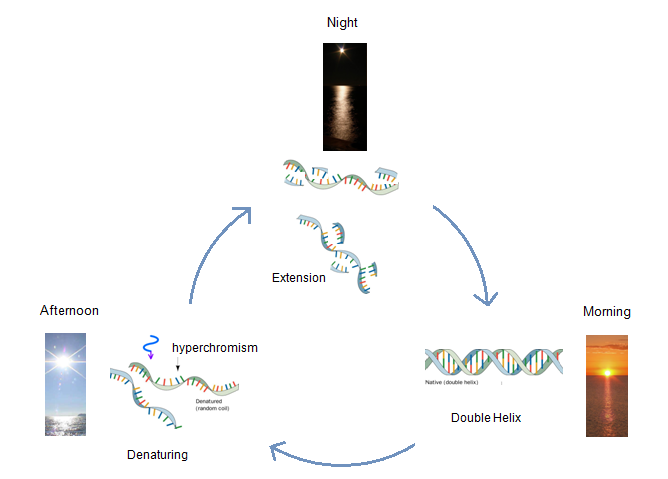} 
\end{center}
\caption{Ultraviolet and Temperature Assisted Reproduction (UVTAR) of RNA and DNA. A proposed mechanism based on photon dissipation for the enzyme-less reproduction of RNA and DNA assisted by the absorption of the prevailing UVC light flux and the high temperatures of the ocean surface during the Archean, including a day/night diurnal warming and cooling cycle of the ocean surface due to the absorption of solar infrared light. Most denaturing would occur in the afternoon when ocean surface temperatures were highest. ``Hyperchromism" refers to the increase (of up to 40\%) in the absorption of photons at UVC wavelenghts ($\sim$260 nm) once RNA or DNA are denatured into single strands. The important aspect of this mechanism is that replication is tied to dissipation, giving proliferation a thermodynamic imperative.}
\label{UVTAR}
\end{figure}

The characteristic sharpness of the temperature denaturing curve for DNA, particularly for larger segments and for acidic pH values $\sim$5 \cite{DubeyTripathi2005} would have facilitated photon-induced denaturing since, once the ambient temperature fell slightly below the melting temperature, only the small amount of energy available in a single photon in the long wavelength UVC region (about 4.8 eV) would have been sufficient to rupture all of the hydrogen bonds between the two strands and separate the strands completely. Double strand RNA has a lower melting temperature and a less sharp denaturing curve. During overnight periods of approximately 7 hours (the Earth rotated more rapidly at the origin of life), the sharp denaturing curve of DNA would mean that the small decrease in ocean surface temperature would have been sufficient to allow for Mg$^{2+}$ mediated extension of the separated single strands \cite{Szostak2012}, completing the reproduction cycle (see figure \ref{UVTAR}).
 
Single strand RNA and DNA is on average 30-40\% more efficient at absorbing and dissipating UVC photons than is double strand RNA or DNA or the hybrid RNA+DNA duplex, an effect known as hyperchromism and is the result of the shadowing or screening of the bases when they are tightly stacked one above the other in a native double strand arrangement \cite{Vekshin2005}. In single strand RNA and DNA, the bases are free to take on arbitrary orientation with respect to the long central axis of the RNA or DNA molecule and so there is less shadowing and therefore greater photon absorption. Denaturing by UVC light would thus increase the rate of dissipation of the solar photon potential in the UVC by about 30-40\% compared to native double strand and, therefore, particular RNA and DNA that could have remained in the denatured state during the day through this mechanism of UVC absorption and dissipation and would have been available for template extension during the night, and thus, in this manner, be ``thermodynamically" selected. If indeed this UVTAR mechanism were operative in the Archean, it would have endowed incipient life with a dissipation-replication relation, which, in fact, remains to this day in living systems and may be useful for placing evolutionary theory on a non-tautological physical-chemical foundation \cite{Michaelian2016}.

Through this UVTAR mechanism, even as the surface temperature of Earth fell below the melting temperature of DNA or RNA, given an incident UVC component in the solar flux available during daylight hours, RNA and DNA oligos could have denatured into single strands without the aid of enzymes. Enzymeless denaturing of RNA at temperatures below its melting temperature is considered to be one of the most difficult problems of the ``RNA World" \cite{Szostak2012}.

Enzymeless photon-induced denaturing is only one half of the complete enzymeless replication problem, the other half being extension (the formation of a new strand using the single strand as a template). However, there exists experimental evidence that extension of a complimentary strand can occur at high temperatures (around 80 $^\circ$C, similar to the surface temperature of the Archean) using chemically activated nucleotides in an aqueous solution containing Mg$^{2+}$ or Zn$^{2+}$ ions \cite{Szostak2012}. There also exists experimental data indicating that enzymeless DNA extension can be speeded up orders of magnitude by including the presence of planer intercalating molecules, such as tryptophan, which act as a kind of glue \cite{HorowitzEtAl2010}. In our proposed UVTAR scenario, nucleotide activation would have been induced by the UVC light during daylight hours and extension would occur overnight when the ocean surface water had cooled by about 3-5 $^\circ$C and there would be no UVC light inhibiting the extension (see figure \ref{UVTAR}). The ocean surface microlayer would have been rich in Mg$^{2+}$ and Zn$^{2+}$ ions and hydrophobic planer intercalating molecules \cite{Michaelian2016}. 
    
Indications from our light-induced denaturing data \cite{MichaelianSantillan2014}, and this previously published data concerning enzymeless extension \cite{Szostak2012}, are that a rudimentary enzymeless RNA or DNA replication driven by the thermodynamic imperative of photon dissipation would have been possible utilizing day-night UVC light cycling, under the physical conditions of Earth's ocean surface during the Archean (high UVC flux and high temperatures with 3-5$^\circ$C diurnal cycling, and pH values of around 6-7).

\section{{\large{}Accumulation of Information through the Dissipation-Replication Mechanism}}

Being an irreversible thermodynamic process, life must have originated, proliferated and even evolved through the dissipation of a generalized chemical potential. At the origin of life in the Archean, we have explicitly identified this chemical potential as the surface UVC photon potential and have proposed an enzymeless ultraviolet and temperature assisted replication mechanism (UVTAR) driven by the dissipation of this light and leading to a dissipation-replication relation \cite{Michaelian2017, Michaelian2016, Michaelian2009, Michaelian2011}.

In the previous sections we have emphasized how the fundamental molecules of life are pigments which absorbed strongly the UVC light available at Earth's surface at the origin of life and throughout the Archean.  We have also emphasized that the amino acids have affinity to RNA and DNA, and, in many cases, particularly to their specific codons or anticodons, evidence for a stereochemical era of association of nucleic acids with amino acids.

In the following subsections we discuss particular characteristics conferred to complexes of  nucleic acids with amino acids which would promote the dissipation-replication relation and lead to greater photon dissipation. We then show how this would have led to the accumulation of information within the nucleic acids relevant to each specific characteristic of the complex promoting photon dissipation.

\subsection{Surface Entrapment / Amphipathic}

Since water absorbs and reflects UVC light (1/e extinction depth of $\sim 1$ m) it would be crucial that photochemical processes be carried out on the surface of Archean oceans where the organic molecules would have been exposed to the greatest UVC photon flux. Under normal circumstances, RNA and DNA would sediment to the ocean floor \cite{MichaelianSantillan2014}, rendering them useless for dissipation.

Eickbush and Moudrianakis \cite{EickbushMoudrianakis1977} found that molecules intercalating RNA and DNA with an alkyl side arm which can fit into the major or minor groove could keep these molecules at the surface. Further investigation led these authors to conclude that even molecules like the polyamine spermidine, which are not intercalating but which can bind to the major or minor groove of DNA or RNA, thereby displacing water and making the molecule more hydrophobic, gives rise to entrapment at the surface. In fact, even the presence of simple mono- or divalent- cations saturating the primary or secondary ionic binding sites of RNA and DNA led to a low level of surface entrapment. Divalent Ca$^{2+}$ and NH$_{4}^{+}$, Ba$^{2+}$ and Na$^+$ ions were found to induce entrapment at minimum concentrations about 30 times lower than that for monovalent ions. As with spermidine, the minimum cation concentration able to support entrapment corresponds to the concentration required to shield the negative phosphate charges \cite{EickbushMoudrianakis1977}.

Corroborating the suggestion that the keeping RNA or DNA at the surface was fundamental to its thermodynamic reason for being of photon dissipation, is the evidence that in the evolutionary sequence of amino acid coding, the first to be codified were most probably the hydrophobic amino acids \cite{WeberLacey1978, CopleyEtAl2005}. The non-polar or hydrophobic amino acids are alanine (Ala), glycine (Gly), valine (Val), leucine (Leu), isoleucine (Ile), proline (Pro), methionine (Met), phenylalanine (Phe), and tryptophan (Trp). According to Trifonov \cite{Trifonov2002}, the early codon table involved only a few aminoacids; Ala, Asp, Gly, Pro, Ser, Thr and Val since these seven amino acids are encoded today by single point mutation derivatives of the presumed earliest parental codon GCU for alanine (Ala), which is a hydrophobic amino acid, as are Gly, Pro and Val\cite{Trifonov2002} (see table \ref{AminoAcids}). There is also a long-recognized relationship between the hydrophobicity of the amino acid and the second base of its codon. Copley et. al \cite{CopleyEtAl2005} suggest that this relationship can be explained if, before the emergence of complex biosynthetic pathways, simple amino acids were synthesized from $\alpha-$keto acid precursors covalently attached to dinucleotides that catalyze the reactions required to synthesize the specific amino acid.

We suggest here that at the very beginnings of life, amphiphatic molecules, having both-+	 hydrophylic and hydrophobic parts, that can adsorb to DNA or RNA and thereby keep them afloat at the surface would have been essential to the photon dissipation program. The amphiphatic amino acids are; lysine, methionine, tryptophan, tyrosine  (see table \ref{AminoAcids}). The amphiphatic amino acids were thus very important to the prevention of sedimentation to the seabed thereby securing the efficacy of the RNA/DNA - amino acid complex in photon dissipation.

\subsection{Charge Neutralization}

It has been suggested that salt concentrations in the Archean would have been 1.5 to 2 times greater than present day ocean levels due to the likelihood that salt entrapment basins on sea shores would have been fewer due to limited formation of continents \cite{Knauth2005} and the fact that bacterial mats preventing salt erosion at these sites \cite{Lovelock1988} would not have existed. At the high ocean surface temperatures of the Archean, particularly near to the equator, neutralization of the negative charges on the RNA or DNA backbone by positively charged amino acids would have allowed overnight extension to occur at higher surface temperatures than otherwise, permitting a UVTAR mechanism to be viable in environments of either higher temperature or lower salt concentration. The amino acids histidine, lysine and arginine have net positive charge and may thus have been important to extension in these environments. RNA or DNA in high nutrient but low salt estuaries could therefore have benefited from such positively charged amino acids. At very high temperatures, positively charged amino acids with their hydrophilic side chains which have affinity to the grooves on RNA or DNA could have performed better at stabilization than salt ions.

Charge neutralization of the phosphates on RNA or DNA could be achieved by molecules which bind ionically and fit within the major or minor grooves, such as spermadine mentioned in the previous subsection. However, over saturation of the sites can lead to precipitation of RNA or DNA so there exists a narrow concentration range of such molecules supporting surface entrapment \cite{EickbushMoudrianakis1977}. Therefore, molecules which were both charge neutralizing and hydrophobic would have performed best at keeping RNA or DNA at the surface and able to template reproduce at the high ocean surface temperatures of the Archean. Of the amphiphatic amino acids (see previous subsection), only lysine is positively charged. In this regard, it is interesting the lysine is highly specifically bonded to its anticodon and is a homocodonic amino acid \cite{WeberLacey1978, HendryEtAl1981a, FoxEtAl1971}.

\subsection{Antenna Molecules}

The nucleobases of RNA and DNA have large molar extinction coefficients for light in the UVC wavelength region and are extremely rapid in the dissipation of their electronically excited singlet states to the ground state, which happens on sub-picosecond time scales
through a conical intersection. Also very rapid ($\sim$2 ps) is the vibrational cooling of the hot molecule to the temperature of its water solvent environment \cite{MiddletonEtAl2009}.

The aromatic amino acids also have strong absorbance in the UVC region, however, their excited state lifetimes are on the order of nanoseconds, or about three orders of magnitude longer than the excited state lifetimes of the nucleic acids and they have a significant quantum efficiency for radiative decay through flourescence \cite{BensassonEtAl1983,CadetVigni1990}. 

The natural stacking affinity that exists between these two classes of aromatic molecules (see section \ref{sec:Affinity}) keeps them close enough to permit excitation energy transfer (EET), allowing the energy of electronic excitation of the amino acid to be dissipated non-radiatively through the conical intersection of the RNA or DNA. Evidence of energy transfer between tryptophan and DNA exists in the observation that the fluorescence of tryptophan is completely quenched when nucleobases or nucleosides are included in solution \cite{MontenayHelene1968,MontenayHelene1971,MontenayEtAl1982}. The amino acid acts as a donor antenna molecule to the acceptor quencher RNA or DNA molecule, increasing the photon dissipation efficacy of the complex over what the components acting separately could achieve. 

The aromatic amino acids are histidine, phenylalanine, tryptophan, and tyrosine. Yarus et al. \cite{YarusEtAl2009} identify these amino acids as among the best candidates for participation in a stereochemical era of genetic code assignments based on amino acid binding at cognate condon or anticodon sites (see table \ref{AminoAcids}).

The binding of aromatic amino acids as antenna molecules to oligonucleotides would play a double role; first that of increasing UVC light dissipation, and secondly, the production of greater local heat to aid in denaturing. Thus, oligos with affinity to aromatic amino acids would have a higher probability of UVTAR replication than random nucleotide sequences with no such affinity. In this manner the affinity between codons and amino acids is preserved by what could be called ``thermodynamic selection'' on dissipation.

\subsection{Intercalation}

Intercalating molecules have been shown to increase the rigidity of RNA and DNA strands, thereby avoiding cyclization which hinders enzymeless extension \cite{KleinwachterKoudelka1964}. In fact, the efficiency for extension in the presence of intercalating molecules can be improved by orders of magnitude \cite{HorowitzEtAl2010}. To fit between the bases, these intercalating molecules must be planer aromatic and have important overlap with the bases for strongest non-covalent binding.

Tryptophan has a large UVC absorption cross section but a slow decay time of nanoseconds and high yield for fluorescence \cite{GudginEtAl1971}. The size of its ring structure is similar to that of the bases so it bonds strongly and can remain attached even to single strand nucleic acid \cite{ToulmeEtAl1974}. Therefore, during the daytime in the Archean it could have acted as an antenna molecule passing its excited state energy to RNA and DNA through resonant energy transfer. This is evidenced by the complete quenching of tryptophan fluorescence when it is mixed with DNA or RNA \cite{RajeswariEtAl1987}.  At night, it could have acted as an extension enhancer. 

Molecular complexes containing a planer moiety are also candidates for intercalation, for example the tripeptide LysTrpLys which intercalates through a two step mechanism; first electrostatic binding between the positively charged Lys residues and the negatively charged phosphate group of the backbone and then stacking of the indole moeity of the Trp residue \cite{RajeswariEtAl1987}. Such a two step binding process has also been observed for other combinations of charged and aromatic amino acids such as LysPheArg  \cite{MondalEtAl2014}. Both of these tripeptides, with confirmed affinity to the corresponding aromatic amino acid codon or anticodon, retain physiological function in contemporary enzymes.

\subsection{Catalysis}

Another amino acid with a strong affinity to its anticodon is histidine \cite{YarusEtAl2009}. This is an 5-member aromatic heterocycle containing two nitrogen atoms and 3 carbon atoms known as an imidazole. Having two conjugate bonds, histidine absorbs strongly in the UVC (212 nm, $\epsilon=5,700$) \cite{SaidelEtAl1952}, however, it has a smaller absorption peak at 280 nm which has been attributed to photon-induced charge transfer (CT) transitions\cite{PrasadEtAl2017}. This is important for the origin of life, because imidazoles have been found to serve as both an acidic and alkaline (pKa = 14.5) catalyst for a very large number of important reactions in contemporary life, for example as a condensing agents for the phosphorylation of the nucleobases and the lipids. Such molecules that can act as both a base and acid catalysts are known as amphoteric.

The stacking of the aromatic rings of histidine with, for example, the imidazole rings of the purine bases, could lead to yet another type of catalysis by, for example, the heat generated from the photon dissipation in histidine catalyzing the thermal chemical reactions in the rest of the purine base production, for example, in the case of the microscopic dissipative structuring of Adenine from HCN \cite{Michaelian2017}.

Other non-aromatic amino acids that have strong absorption in the UVC Archean atmospheric window region attributed to charge transfer transitions, are the charged amino acids Lys, Glu monosodium salt (Glu$\cdot$Na), Arg, and Asp potassium salt (Asp$\cdot$K) \cite{PrasadEtAl2017}. According to Yarus et al. \cite{YarusEtAl2009} these, along with histidine, have strong association with either their cognate condons or anticodons. These, therefore, could have acted as catalysts for early biotic photochemistry.

\section{Dissipation-Replication and Amino Acid Affinity to Codon}
In table \ref{AminoAcids} we compare the amino acids with the above listed characteristics which would have promoted the dissipation-replication relation through the UVTAR mechanism with the the strength of association between an amino acid and their cognate condons or anticodons as determined by Yarus and coworkers \cite{YarusEtAl2009}. It is striking that those amino acids with characteristics that could enhance the dissipation-replication relation are just those with strong association with their codons or anticodons. 

We suggest that these characteristics are therefore the basis of the specificity of the interaction between nucleic acids and amino acids.  Optimizing the dissipation of the long wavelength UVC region of the Archean solar spectrum could thus have been the origin of the information encoding in RNA and DNA. This is still relevant for the genomes of today's organisms which are essentially nothing more than blueprints for the construction of biosynthetic pathways to dissipate the external generalized chemical potentials existent (or once existent) in the organisms environment.  
\newpage
\begin{table}[h]
\begin{centering} \tiny
\begin{tabular}{|c|c|c|c|c|c|c|c|c|}
\hline 
\multirow{2}{*}{\textbf{Amino acid}} & \multirow{2}{*}{\textbf{Abbreviation}} & \multirow{2}{*}{\textbf{Codon}} & \textbf{Codon/Anticodon}  &\multirow{2}{*}{\textbf{Amphipathic}} & \textbf{Antenna} & \multirow{2}{*}{\textbf{Intercalating}} & \multirow{2}{*}{\textbf{Catalysis}} & \textbf{Charge}\tabularnewline
 &  & & \textbf{Affinity} &  & \textbf{260 nm} & & & \textbf{Neutralizing}\tabularnewline
\hline 
\multicolumn{9}{|c|}{Aliphatic not polar R group   (Hydrophobics)} \tabularnewline
\hline 
\multirow{4}{*}{Glycine} & \multirow{4}{*}{Gly} & GGU &  &  & \multirow{4}{*}{} & & &\tabularnewline
 & & GGC & & & & & & \tabularnewline
 & & GGA & & & & & & \tabularnewline
 & & GGG & & & & & & \tabularnewline
\hline 
\multirow{4}{*}{Alanine} & \multirow{4}{*}{Ala} & GCU &  & & \multirow{4}{*}{} & & & \tabularnewline
 & & GCC & & & & & & \tabularnewline
 & & GCA & & & & & & \tabularnewline 
 & & GCG & & & & & & \tabularnewline 
\hline 
\multirow{4}{*}{Proline} & \multirow{4}{*}{Pro} & CCU &  & & \multirow{4}{*}{} & & &\tabularnewline
& & CCC & & & & & & \tabularnewline
& & CCA & & & & & & \tabularnewline
& & CCG & & & & & & \tabularnewline 
\hline 
\multirow{4}{*}{Valine} & \multirow{4}{*}{Val} & GUU &  & & \multirow{4}{*}{} & & &\tabularnewline
& & GUC & & & & & & \tabularnewline
& & GUA & & & & & & \tabularnewline
& & GUG & & & & & & \tabularnewline 
\hline 
\multirow{6}{*}{Leucine} & \multirow{6}{*}{Leu} & UUA & & & \multirow{6}{*}{} & & &\tabularnewline
& & UUG & & & & & & \tabularnewline
& & CUU & & & & & & \tabularnewline
& & CUC & & & & & & \tabularnewline
& & CUA & & & & & & \tabularnewline
& & CUG & & & & & & \tabularnewline
\hline 
\multirow{3}{*}{Isoleucine} & \multirow{3}{*}{Ile} & AUU & s/ & & \multirow{3}{*}{yes} & & & \tabularnewline
& & AUC & & & & & & \tabularnewline
& & AUA & /s & & & & & \tabularnewline
\hline 
Metionine & Met & AUG & & yes & yes & & &\tabularnewline
\hline 
\multicolumn{9}{|c|}{Aromatic R group (Slightly hydrophobic)}\tabularnewline
\hline 
\multirow{2}{*}{Phenylalanine} & \multirow{2}{*}{Phe} & UUU &  &  & \multirow{2}{*}{yes} &
\multirow{2}{*}{yes} &  & \tabularnewline
& & UUC & /m & & & & & \tabularnewline
\hline 
\multirow{2}{*}{Tyrosine} & \multirow{2}{*}{Tyr} & UAU & /s & \multirow{2}{*}{yes} & \multirow{2}{*}{yes} & \multirow{2}{*}{yes} & & \tabularnewline
& & UAC & /w &  & &  & &\tabularnewline
\hline 
Tryptophan & Trp & UGG & /s & yes & yes & yes & & \tabularnewline
\hline
\multicolumn{9}{|c|}{Polar R group without charge}\tabularnewline
\hline 
\multirow{6}{*}{Serine} & \multirow{6}{*}{Ser} & UCU & &   & \multirow{6}{*}{} & & & \tabularnewline
& & UCC & & & & & & \tabularnewline
& & UCA & & & & & & \tabularnewline
& & UCG & & & & & & \tabularnewline
& & AGU & & & & & & \tabularnewline
& & AGC & & & & & & \tabularnewline
\hline 
\multirow{4}{*}{Threonine} & \multirow{4}{*}{Thr} & ACU & & & \multirow{4}{*}{} & & & \tabularnewline
& & ACC & & & & & & \tabularnewline
& & ACA & & & & & & \tabularnewline
& & ACG & & & & & & \tabularnewline
\hline 
\multirow{2}{*}{Cysteine} & \multirow{2}{*}{Cys} & UGU &  &  & \multirow{2}{*}{} & & & \tabularnewline
& & UGC & & & & & & \tabularnewline
\hline 
\multirow{2}{*}{Aspargine} & \multirow{2}{*}{Asn} & AAU &  &  & \multirow{2}{*}{} & & & \tabularnewline
& & AAC & & & & & & \tabularnewline
\hline 
\multirow{2}{*}{Glutamine} & \multirow{2}{*}{Gln} & CAA &  &  & & & & \tabularnewline
& & CAG & & & & & & \tabularnewline
\hline 
\multicolumn{9}{|c|}{R group positively charged}\tabularnewline
\hline 
\multirow{2}{*}{Lysine} & \multirow{2}{*}{Lys} & AAA & /s & \multirow{2}{*}{yes} & \multirow{2}{*}{yes$^{*}$} &  & \multirow{2}{*}{yes} & \multirow{2}{*}{yes}\tabularnewline 
 & & AAG &  &  &  & & & \tabularnewline
\hline 
\multirow{2}{*}{Histidine} & \multirow{2}{*}{His} & CAU & /w & & 
\multirow{2}{*}{yes$^{*}$} & \multirow{2}{*}{yes} & \multirow{2}{*}{yes} & \multirow{2}{*}{yes} \tabularnewline
 & & CAC & /s &  &  & & & \tabularnewline
\hline 
\multirow{5}{*}{Arginine} & \multirow{5}{*}{Arg} & CGU &  &  & \multirow{5}{*}{yes$^{*}$} & & \multirow{5}{*}{yes} & \multirow{5}{*}{yes} \tabularnewline
& & CGC &  &  &  & & & \tabularnewline
& & CGA & /s  &  &  & & & \tabularnewline
& & CGG & w/ &  &  & &  & \tabularnewline
& & AGA &  &  &  & & & \tabularnewline
& & AGG & s/ &  &  & &  &\tabularnewline
\hline 
\multicolumn{9}{|c|}{R group negatively charged}\tabularnewline
\hline 
\multirow{2}{*}{Aspartic acid} & \multirow{2}{*}{Asp} & GAU &  &   & \multirow{2}{*}{yes$^{*}$} & & \multirow{2}{*}{yes} & \tabularnewline
& & GAC &  &  &  & &  & \tabularnewline
\hline 
\multirow{2}{*}{Glutamic acid} & \multirow{2}{*}{Glu} & GAA &  &  & \multirow{2}{*}{yes$^{*}$} & & \multirow{2}{*}{yes} & \tabularnewline
& & GAG &  &  &  & & & \tabularnewline
\hline
\end{tabular}
\par\end{centering}

\caption{Attributes of the common amino acids classified by hydrophobicity and by electrostatic charge. Those amino acids having a strong (s; probability for non-specificity, Corr $P <10^{-4}$) or moderate (m; Corr $P<10^{-3}$ )  or weak (w; Corr $P<10^{-2}$) association of their binding sites with their cognate codons/anticodons (Corr $P=1$ implies no specificity; see table 1 of \cite{YarusEtAl2009}) are also those that have the greatest number of characteristics relevant to a possible photon dissipation-replication relation at the origin of life. The ``$*$'' indicates amino acids that absorb in the 260 nm window through charge transfer (CT) transitions \cite{PrasadEtAl2017}. For the ultraviolet spectra of $\alpha-$amino acids. see \cite{Wetlaufer1963} \label{AminoAcids} }
\end{table}

\section{Discussion and Conclusions}

We have suggested here that the first information programmed into the microscopic dissipative structures known as RNA and DNA had a direct thermodynamic utility; that of providing a physical scaffolding for the attachment of molecules which aid in the dissipation of the prevailing UVC solar photon potential. We postulate that this was achieved through a dissipation-replication relation for RNA and DNA employing an ultraviolet and temperature assisted mechanism for enzymeless replication (UVTAR) as presented above. Since this mechanism relates replication with UVC photon dissipation, it provides a non-equilibrium thermodynamic and physical-chemical basis for the accumulation of information relevant to the origin and evolution of life.

The properties of amino acids which may have been important to enhancing dissipation are; their amphiphatic character required for entrapment of RNA and DNA at the ocean surface, their UVC light antenna properties for greater photon dissipation and local heating to foment denaturation, intercalating facility to enhance overnight extension and reduce photoproducts, charge neutralizing properties to foment extension at earlier higher temperatures, and catalytic properties (e.g. charge transfer) to foment proliferation.

Yarus and coworkers \cite{YarusEtAl2009} building on the work of others \cite{WeberLacey1978, HendryEtAl1981a, FoxEtAl1971} have unequivocally demonstrated that there exists a chemical affinity for a subset of amino acids to sites containing their cognate codons or anticodons, and that this is indicative of a stereochemical era near the beginning of life. Here we have shown that it is exactly these amino acids having affinity to their codons or anticodons that have the properties necessary for fomenting an efficient dissipation through the UVTAR mechanism for RNA and DNA enzymeless replication. The data of Yarus and coworkers was obtained with RNA, however the codon scaffolding is very similar for DNA.

According to our analysis, the first amino acids to be encoded were the aromatic (phenylalanine, tyrosine, tryptophan) and the positively charged (lysine, histidine, arginine) groups which are those of the 20 commonly used by life with the greatest number of characteristics most beneficial to the UVTAR of RNA and DNA and these have strong association with their cognate codons and anticodons (see table \ref{AminoAcids}). Since the greatest number (8), and indeed the strongest, of associations are found among the anticodons rather than the codons, for which only 3 associations were found (see table \ref{AminoAcids}), and, in fact, there are no amino acids associated only with their codons \cite{YarusEtAl2009}, we conclude that most of the earliest evolution of life probably involved DNA rather than RNA, thus challenging the RNA World hypothesis. Other evidence that would favor at least a contemporary RNA and DNA World would be; i) there appears to be more direct stereochemical relationships between amino acid R groups and the purines and pyrimidines of DNA than of RNA \cite{HendryEtAl1981a}, ii) the high temperatures of the ocean surface during the Archean would make double helix DNA less prone to hydrolysis and therefore significantly more stable than RNA, and iii) from the perspective of the thermodynamic dissipation theory of the origin of life, there is little difference between RNA and DNA in their UVC photon dissipation capacity. 

It is sometimes argued that the aromatic amino acids must have appeared later in the evolutionary history of life due to their contemporary complex biosynthetic pathways. However, this ignores the relatively simple UV photochemical pathways to such molecules which could have existed throughout the whole of the Archean. For example, there are relatively few steps involved in the production of the even more complex nucleobase adenine starting from HCN in water \cite{Michaelian2017}. If adenine was produced abiotically early in the history of life, then it is probable that the aromatic amino acids could also have been. Indeed, aromatic hydrocarbons have been found in meteorites \cite{BottaEtAl2008} and interstellar space \cite{MichaelianSimeonov2017, AllamandolaEtAl1985}. The stereochemical affinity of these to RNA and DNA \cite{YarusEtAl2009} and the large number of characteristics useful for the proposed enzymeless replication of RNA and DNA (table \ref{AminoAcids}) argue for these being among the first. 

It is also been argued by Copley et al. \cite{CopleyEtAl2005} that the present triplet code originated from a binary one in which the first two bases of the codon specified uniquely 14 of the 22 bases used by life and that this was related to the possibility of dinucleotides acting as catalysts for the production of the most simple amino acids from $\alpha$-keto acids. Although this hypothesis is appealing, and indeed we have suggested that intermediates on the way to the purines, such as imidazoles could have acted as photochemical catalysts for other fundamental molecules \cite{Michaelian2017}, there is no evidence for a stereochemical affinity between the dinucleotides and these suggested amino acids or their precursor $\alpha$-keto acids, except, perhaps for the negatively charged amino acids Glu and Asp. Furthermore, the contemporary chemical route to $\alpha$-keto acids is quite complex and these do not appear to have a photochemical origin. We therefore suggest that the associations suggested by Copley et al. may indeed have been some of the first steps on route to a purely chemical (independent of UVC photons) biosynthetic pathway of the simple amino acids, but that this probably appeared much later in the evolutionary history of life, i.e. much after the appearance of stereochemical associations (as found by Yarus et al.) which we suggest would have been important for direct dissipation of the long wavelength UVC region.

Amino acids without properties relevant to the UVTAR process must have been relevant to reproduction at some point in the history of life and were probably assigned codons once metabolic pathways had been established in a more advanced coevolutionary process \cite{Woese1966, Wong1975}. During coevolution, not only could amino acids have been added to the code, but earlier codonic assignments may have been changed. Despite this, it is important to emphasize that there are sequences of nucleotides and its codonic assignments that have remained immutable since ancestral times \cite{Rossmann1974, DayhoffEtAl1978}.

Finally, we end with a quote by Carl Woese (1967) reproduced in the seminal paper of Yarus et al. \cite{YarusEtAl2009} cited frequently above;
\begin{quote} ``I am particularly struck by the difficulty of getting [the genetic code] started unless there is some basis in the specificity of interaction between nucleic acids and amino acids or polypeptide to build upon.'' \end{quote} 
We hope that we have presented adequately the work of Yarus and coworkers and others demonstrating the specificity in the interaction between nucleic acids and amino acids. Our work presented here is a proposal for the the basis of that specificity; that of increasing photon dissipation efficacy of the RNA/DNA-amino acid complex which leads to higher differential replication through a UVTAR mechanism operating under non-equilibrium thermodynamic imperatives.\\
\ \\
{\noindent \bf \large Acknowledgements:}
The authors gratefully acknowledge the financial support of DGAPA-UNAM, project number IN102316, and the National Quality Postgraduate Program CONACyT.

%


\end{document}